\documentclass[preprint,floats,aps, superscriptaddress,nofootinbib,floatfix,showkeys]{revtex4-2}
\usepackage{times} 
\usepackage{amssymb,amsmath}
\usepackage{csquotes}
\usepackage{physics}
\usepackage{bm}
\usepackage{nicefrac}
\usepackage{mathtools}
\usepackage{setspace}
\usepackage{booktabs}
\usepackage{upgreek}
\usepackage[per-mode=fraction]{siunitx}
\usepackage{float}

\usepackage[usenames,dvipsnames]{color}
\usepackage[pdftex]{epstopdf}
\usepackage[pdftex]{graphicx}
\usepackage{hyperref}
\hypersetup{pdftitle={Manuscript on } 
    pdfauthor={Justus Niehoff, Florian Voss and Uwe Thiele},  
pdfproducer={lateX},
pdfview=FitV,       
pdfstartview=FitB,
linkcolor=blue,     
citecolor=blue,     
urlcolor=blue,      
breaklinks=true,    
colorlinks=true,
citebordercolor=0 0 0,  
filebordercolor=0 0 0,
linkbordercolor=0 0 0,
menubordercolor=0 0 0,
urlbordercolor=0 0 0,
pdfhighlight=/I,
pdfborder=0 0 0,   
bookmarksopen=true,
bookmarksnumbered=true
}
\usepackage{tcolorbox}
\DeclareGraphicsExtensions{.jpg, .pdf, .tif, .png}
\usepackage[usenames,dvipsnames]{color}
\usepackage[normalem]{ulem}

\newcommand{\subref}[2]{\hyperref[#1]{\ref{#1}#2}}

\newcommand{\vecg}[1]{\boldsymbol{#1}}

\newcommand{\tensg}[1]{\boldsymbol{\underline{#1}}}

\usepackage{booktabs}
\usepackage{multirow}
\makeatletter
\renewcommand{\p@subsection}{}
\renewcommand{\p@subsubsection}{}
\renewcommand{\@seccntformat}[1]{\csname the#1\endcsname\quad}
\makeatother

\begin{document}
\count\footins = 1000 

\title{Gradient dynamics model for chemically driven running drops}
\author{Justus Niehoff}
\email{\textcolor{blue}{jniehoff@uni-muenster.de}}
\thanks{ORCID ID: 0009-0006-1533-7062}
\thanks{JN and FV contributed equally to this work.}
\affiliation{Institute of Theoretical Physics, University of M\"unster, Wilhelm-Klemm-Str.\ 9, 48149 M\"unster, Germany}

\author{Florian Voss}
\email{\textcolor{blue}{f\_voss09@uni-muenster.de}}
\thanks{ORCID ID: 0009-0003-9679-035X}
\thanks{JN and FV contributed equally to this work.}
\affiliation{Institute of Theoretical Physics, University of M\"unster, Wilhelm-Klemm-Str.\ 9, 48149 M\"unster, Germany}

\author{Uwe Thiele}
\email{\textcolor{blue}{u.thiele@uni-muenster.de}}
\homepage{http://www.uwethiele.de}
\thanks{ORCID ID: 0000-0001-7989-9271}
\affiliation{Institute of Theoretical Physics, University of M\"unster, Wilhelm-Klemm-Str.\ 9, 48149 M\"unster, Germany}
\affiliation{Center for Data Science and Complexity (CDSC), University of M\"unster, Corrensstr.\ 2, 48149 M\"unster, Germany}

\begin{tcolorbox}[title=Publication note,
title filled=false,
colback=red!5!white,
colframe=red!75!black,sharp corners]
This version of the article has been accepted for publication after peer-review. It has been modified to account for most post-acceptance improvements and corrections, but it is not the version of record. The version of record is available at:
\\
\\
J.~Niehoff, F.~Voss and U.~Thiele.
\newblock Gradient dynamics model for chemically driven running drops.
\newblock \emph{Eur. Phys. J. Spec. Top.}, 2026{\natexlab{a}}.
\newblock \doi{10.1140/epjs/s11734-026-02296-w}.
\end{tcolorbox}

\begin{abstract}
We present a thermodynamically consistent model for chemically driven running drops on a solid substrate with reversible substrate adsorption of a wettability-changing chemical species. We consider drops confined to a vertical gap, thereby allowing us to first obtain a gradient dynamics description of the closed system, corresponding to a set of coupled dynamical equations for the drop profile and the chemical concentration profiles of species on the substrate and in both fluids (drop, ambient medium). Chemostatting the species in the drop and the ambient medium, we then derive a reduced model for the dynamics of the drop and the adsorbate on the substrate. When the externally imposed chemical potentials are distinct, the system is driven away from thermodynamic equilibrium, allowing for sustained drop self-propulsion across the substrate due to a wettability contrast maintained by chemical reactions. We numerically study the resulting running drops and show how they emerge from drift-pitchfork bifurcations. 
\end{abstract}

\maketitle
\section{Introduction}
It is well-known that external gradients can drive fluid motion,~e.g., causing liquid drops to run uphill on a chemically altered substrate \citep{ChWh1992s}, and driving the downwards motion of bubbles in a vertically increasing temperature field \citep{YoGB1959jofm}. By contrast, insights into self-induced drop and colloidal particle motion are relatively recent. Examples include catalytic Janus particles \citep{GoLA2005prl}, enzyme-enriched biomolecular condensates \citep{DGMF2023prl,GDMF2024prr}, active polar drops \citep{TSJT2020pre,StJT2022sm} and droplet microswimmers \citep{Mich2023arfm,SeFM2016epjt}. Often, such systems are also considered as particular instances of active matter, where self-organized local energy influxes prevent relaxation to thermodynamic equilibrium and drive sustained dynamic behavior,~e.g., self-propulsion. A classical example for self-propelled liquid drops on solid substrates are chemically driven running drops \citep{DoOn1995prl,SMHY2005prl,LeLa2000jacs,LeKL2002pre, ThJB2004prl,JoBT2005epje,SKYN2005pre}, where drops are laden with a substance that can reduce (or generally alter) the wettability of the underlying substrate via chemical reactions. Initially stationary drops then sit on a less wettable \enquote{patch} due to the occurring reactions. Their spatial symmetry is eventually broken by fluctuations and they are driven away from their current position on the substrate. The motion is then further maintained by the reactions that continuously render the underlying substrate less wettable, thereby sustaining a driving wettability contrast. A distinction is made between scenarios where the substrate is either irreversibly \citep{DoOn1995prl,LeLa2000jacs, LeKL2002pre} or reversibly \citep{SMHY2005prl,SKYN2005pre} altered. In the former case, drops avoid their own less wettable \enquote{trail}, whereas in the latter case, drop motion is unrestricted. Reversible substrate coverage occurs,~e.g., when the adsorbed chemical dissolves into an ambient fluid once the drop has passed the corresponding area on the substrate. 

In both cases, the experimentally observed dynamics has been modeled by combining a thin-film description of partially wetting drops with reaction-diffusion dynamics of the chemical species on the substrate \citep{ThJB2004prl,JoBT2005epje}. In these models, the chemomechanical coupling occurs via a concentration-dependent disjoining (or Derjaguin) pressure that describes wettability. Notably, it is assumed that the coupling is \enquote{one-sided},~i.e., it affects neither diffusion on the substrate nor the reaction kinetics, which is assumed of ideal mass-action type and only occurs below the drop. These strong assumptions imply that this simple approach is not thermodynamically consistent.

 A thermodynamically consistent model for the \textit{irreversible} case is proposed in Ref.~\citep{VoTh2024jem}. There, reciprocal couplings between, on the one hand, a concentration-dependent disjoining pressure and, on the other hand, the reaction kinetics and diffusion follow directly from the thermodynamic structure of the model. The corresponding back-couplings yield modifications of diffusion and reaction kinetics in the contact line region, resulting,~e.g., in concentration gradients even at thermodynamic equilibrium. By treating the drop bulk as a large particle reservoir with an externally imposed chemical potential (chemostat) for the chemical species, drops may then self-propel across the substrate. However, without nontrivial flows imposed at the system boundaries, this self-propulsion is transient and drops ultimately come to rest. This is either because they are slowed down by the continuous diffusion of the chemical in their trail and due to concentration gradients forming at the advancing contact line (caused by the reciprocal coupling of wettability and reaction kinetics), or because they run into their own \enquote{trail}. This process always reflects a relaxation toward thermodynamic equilibrium, independently of the value of the externally imposed chemical potential, and is characterized by a monotonic decrease of the semi-grand free energy \citep{AAFE2024jcp} in contrast to the increasing free energy shown in Ref.~\citep{VoTh2024jem}. Note that for a domain of infinite extent the relaxation process can take the form of a steadily moving drop in analogy of a drop sliding down an infinitely extended incline.

Here, we introduce a thermodynamically consistent (gradient dynamics) model for the \textit{reversible} case. To this end, we consider sessile drops in an ambient medium that is confined in a vertical gap between two parallel plates. For the closed system, this geometry allows us to derive a gradient dynamics description for the drop and the concentrations of the chemical substance in the drop, on the substrate and in the ambient fluid. Note that this is in contrast to the description of the reversible case in Ref.~\citep{VoTh2024jem}, where no such vertical confinement is necessary to capture the dynamics of the closed system. By then treating the drop and the surrounding medium as chemostats, respectively representing a large particle source and sink, we obtain a reduced model for the drop and the adsorbate on the substrate. This reduced model effectively differs from a similarly reduced model for the reversible case introduced in \citep{VoTh2024jem} by an additional reaction term that captures the exchange with the ambient fluid. When the externally imposed chemical potentials of the chemostats are distinct, conditions for thermodynamic equilibrium are violated and the system is driven out of equilibrium. For sufficiently strong nonequilibrium driving, drops then exhibit self-propulsion across the substrate that is sustained indefinitely. This is in fundamental contrast to the case of a single chemostat \citep{VoTh2024jem},~i.e., irreversible substrate coverage, which is always relaxational.

The outline of this paper is as follows. In Sec.~\ref{sec:gradient_dynamics}, we briefly review the general gradient dynamics structure for spatially extended systems, including chemical reactions. We refer to Ref.~\citep{GrootMazur1984} for a general account of nonequilibrium thermodynamics. Specifically for chemical reactions, see Refs.~\citep{RaEs2016prx, APFE2021jcp, AAFE2024jcp}. For applications to (reactive) thin-film hydrodynamics, see Refs.~\citep{Thie2018csa,VoTh2024jem, VoTh2025prf,VoTh2025arxiv}. In Sec.~\ref{sec:running_drops_model}, we present our model based on a gradient dynamics structure. We first provide a description of the relaxational closed system and then introduce nonequilibrium driving by chemostatting reactants in the drop and the ambient medium. The resulting persistent drop self-propulsion is numerically studied in some detail. In Sec.~\ref{sec:conclusion}, we conclude and discuss possible extensions of the model.

\section{N-field gradient dynamics model with chemical reactions}\label{sec:gradient_dynamics}
We consider a closed isothermal system of $N$ scalar field variables $\vecg{\psi} = (\psi_1, \ldots, \psi_N)^\text{T}$. These fields are functions of time $t$ and position $\vec{x}$ and may represent profiles of chemical concentrations, film thicknesses or mechanical deformations. We assume that the system is overdamped,~i.e., inertial effects are negligible. For systems sufficiently close to thermodynamic equilibrium, the time evolution is then described by equations in gradient dynamics form \citep{Thie2018csa, VoTh2024jem}
\begin{equation}
\partial_t \vecg{\psi} \left(\vec{x},t\right) = \vec{\nabla}\cdot\left[\tensg{\mathcal{Q}}\left(\vecg{\psi}\right)\vec{\nabla}\frac{\delta\mathcal{F}}{\delta\vecg{\psi}}\right]-\tensg{\mathcal{M}}\left(\vecg{\psi}\right)\frac{\delta\mathcal{F}}{\delta\vecg{\psi}}+\vecg{\mathcal{R}}\left(\vecg{\psi}\right), \label{eq:gradient_dynamics_general}
\end{equation}
where $\mathcal{F}$ is the free energy functional of the system. The partial derivative with respect to time is denoted by $\partial_t$ and $\vec{\nabla}$ is the spatial gradient operator. The variational derivatives $\delta\mathcal{F}/\delta\vecg{\psi} = \left(\delta\mathcal{F}/\delta\psi_1,\ldots, \delta\mathcal{F}/\delta\psi_N\right)^\text{T}$ represent thermodynamically conjugate quantities to the fields such as pressures or chemical potentials. Equation (\ref{eq:gradient_dynamics_general}) comprises three contributions. The first term corresponds to spatial transport, driven by,~e.g., spatial gradients of variations. The second term represents local exchanges between the fields,~e.g., due to evaporation or imbibition. There, the currents are driven by local differences between variations. Both mobility matrices, $\tensg{\mathcal{Q}}$ and $\tensg{\mathcal{M}}$, are symmetric and positive semi-definite, reflecting microscopic reversibility and a nonnegative entropy production \citep{GrootMazur1984}. The last term represents contributions from chemical reactions and has the form $\vecg{\mathcal{R}} = -\sum_r\left(\vecg{\nu}^+_r-\vecg{\nu}^-_r\right)j_r$, where $\vecg{\nu}^{+}_r$ and $ \vecg{\nu}^-_r$ are the (tuples of) stoichiometric coefficients in the forward and backward direction of reaction $r$, respectively. We assume that in total $R$ reactions take place. The entries $\nu^+_{r\alpha}$ and $\nu^-_{r \alpha}$ with $\alpha=1, \ldots, N$ and $r=1, \ldots, R$ reflect the stoichiometry of the field $\psi_\alpha$ in reaction $r$. The corresponding reaction current $j_r= j^+_r-j^-_r$ similarly contains contributions from the forward ($j_r^+$) and backward ($j_r^-$) directions. Note that $j_r^+, j_r^-\geq 0$ and every reaction is assumed reversible. Further, the forward and backward currents of each reaction obey local detailed balance \citep{APFE2021jcp,AAFE2024jcp}
\begin{equation}
k_\mathrm{B} T \ln \frac{j^+_r}{j^-_r}  = \langle\vecg{\nu}^+_r-\vecg{\nu}^-_r, \frac{\delta\mathcal{F}}{\delta\vecg{\psi}}\rangle, \label{eq:local_detailed_balance}
\end{equation}
where $\langle \vecg{\varphi}, \vecg{\omega}\rangle = \sum_{\alpha=1}^N \varphi_\alpha \omega_\alpha$.\footnote{We remark that in Eq.~(\ref{eq:local_detailed_balance}), we have not restricted ourselves to fields $\psi_a$ that actually correspond to chemical concentrations. However, we assume that all fields that engage in some chemical reaction can be linearly transformed to some effective chemical concentration or density. This applies,~e.g., to film thicknesses of homogeneous liquids.} Equation (\ref{eq:local_detailed_balance}) is equivalent to the general form 
\begin{equation}
j_r = k_r\left[\exp\left(\frac{1}{k_\mathrm{B} T}\langle\vecg{\nu}^+_r, \frac{\delta\mathcal{F}}{\delta\vecg{\psi}}\rangle\right)-\exp\left(\frac{1}{k_\mathrm{B} T}\langle\vecg{\nu}^-_r, \frac{\delta\mathcal{F}}{\delta\vecg{\psi}}\rangle\right)\right], \label{eq:local_detailed_balance_equiv}
\end{equation}
where $k_r\geq 0$ is an unspecified rate function that may,~e.g., depend on the fields. Note that Eq.~(\ref{eq:local_detailed_balance}) is a nonlinear flux-force relation that connects the ratio of the forward and backward reaction currents to the chemical affinity $\langle\vecg{\nu}^+_r-\vecg{\nu}^-_r, \frac{\delta\mathcal{F}}{\delta\vecg{\psi}}\rangle$ \citep{GrootMazur1984},~i.e., the thermodynamic force that drives the reaction. By contrast, the other fluxes in Eq.~(\ref{eq:gradient_dynamics_general}) are \textit{linear} superpositions of (spatial gradients of) the variations, with coefficients given by $\tensg{\mathcal{Q}}$ and $\tensg{\mathcal{M}}$. We note that Eq.~(\ref{eq:local_detailed_balance_equiv}) may be linearized in the chemical affinity to yield a flux that is linear in the variations and consistent with linear nonequilibrium thermodynamics \citep{GrootMazur1984}. That is, close to thermodynamic equilibrium, $\vecg{\mathcal{R}}$ may be reduced to an expression of the form of the second term in Eq.~(\ref{eq:gradient_dynamics_general}). The opposite direction,~i.e., the extension of the common linear descriptions of,~e.g., evaporation and condensation into the nonlinear regime is an interesting open question.

From the structure of Eq.~(\ref{eq:gradient_dynamics_general}) as well as from Eq.~(\ref{eq:local_detailed_balance}) and the positive semidefiniteness of $\tensg{\mathcal{Q}}$ and $\tensg{\mathcal{M}}$, it follows that ${\text{d}\mathcal{F}}/\text{d}t\leq 0$ \citep{AAFE2024jcp, VoTh2024jem},~i.e., the system monotonically relaxes to thermodynamic equilibrium (or, since Eq.~(\ref{eq:gradient_dynamics_general}) does not include noise, to some metastable steady state). In particular, this precludes time-periodic behavior such as sustained oscillations\footnote{Because of the symmetry of $\tensg{\mathcal{Q}}$ and $\tensg{\mathcal{M}}$, damped oscillations are also excluded.} and traveling states, as well as spatio-temporal chaos, as commonly observed in systems that are driven away from thermodynamic equilibrium \citep{VoTh2025prf, GLFT2025prl,FrTh2023prl,BrMa2024prx,SaAG2020prx,FrWT2021pre,YoBM2020pnasusa}. Sources for nonequilibrium driving may be introduced in Eq.~(\ref{eq:gradient_dynamics_general}) in several ways. One could,~e.g., directly add nonvariational terms, reflecting some internal forcing, and thereby break the structure of Eq.~(\ref{eq:gradient_dynamics_general}) \citep{TSJT2020pre, SaAG2020prx, FrWT2021pre}. Alternatively, one could break the thermodynamic constraints imposed on $\tensg{\mathcal{Q}}$ or $\tensg{\mathcal{M}}$ by rendering them indefinite\footnote{In some cases, this is equivalent to the addition of nonvariational terms \citep{FHKG2023pre, GLFT2025prl}.} \citep{FHKG2023pre,GLFT2025prl}, or break (local) detailed balance (\ref{eq:local_detailed_balance}) \citep{AAFE2024jcp}. The latter may be realized via \textit{chemostatting},~i.e., particle exchange with external reservoirs. Then, the chemical potentials of some of the reacting species are imposed externally and appear as additional terms (that are not related to the variations) on the right hand side of Eq.~(\ref{eq:local_detailed_balance}). Nonequilibrium driving then arises from a violation of thermodynamic equilibrium conditions,~e.g., when the imposed values of several chemical potentials are mutually incompatible at equilibrium or when spatial gradients in chemical potentials are imposed \citep{AAFE2024jcp}. Often, the presence of chemostats is rationalized by assuming that some species are highly abundant and their concentrations then remain approximately constant. For closed finite systems, it can be shown under certain restrictions that when some species become highly abundant, they may indeed be treated as chemostats and the system behaves like an open one \citep{ReEA2025jcp}.

To obtain specific (passive) models, Eq.~(\ref{eq:gradient_dynamics_general}) is supplemented with expressions for the free energy $\mathcal{F}$, the mobilities $\tensg{\mathcal{Q}},\tensg{\mathcal{M}}$, the stoichiometries $\vecg{\nu}^+_r, \vecg{\nu}^-_r$ and the reaction rate functions $k_r$. Examples that are then captured by Eq.~(\ref{eq:gradient_dynamics_general}) include diffusion equations \citep{ThAP2016prf,VoTh2024jem, Doi2011jpcm}, Cahn-Hilliard models \citep{Doi2011jpcm,Bray1994ap, Zwic2022cocis}, (non-ideal) reaction-diffusion systems \citep{AAFE2024jcp}, reactive \citep{VoTh2024jem,VoTh2025prf, VoTh2025arxiv} and non-reactive \citep{ThTL2013prl, ThAP2012pf,ThAP2016prf, HeST2021sm, HDGT2024l, DiTh2025prf} thin-film models and other soft matter systems \citep{Doi2011jpcm}.

\section{Chemically driven running drops}\label{sec:running_drops_model}
\begin{figure}
\centering
\includegraphics[scale=0.45]{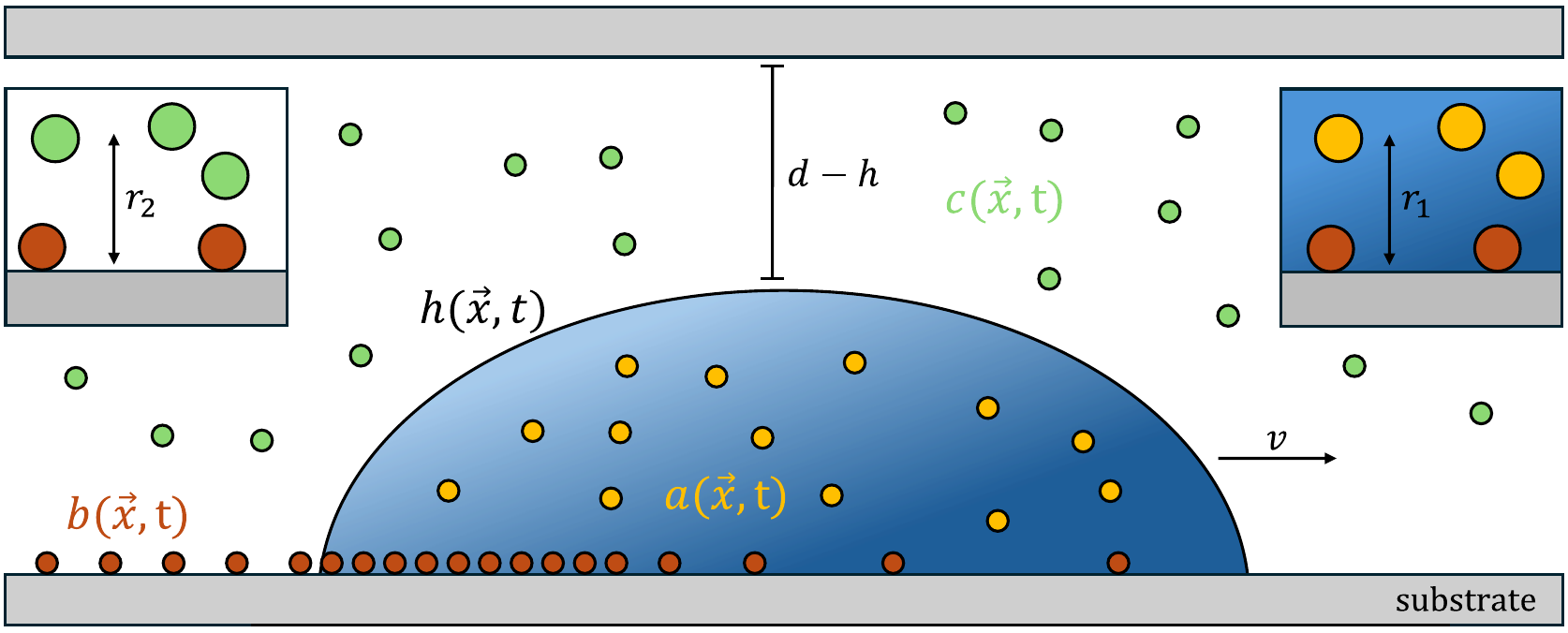}
\caption{Sketch of a chemically driven running drop confined to a vertical gap of height $d$. The local film height is denoted by $h(\vec{x}, t)$. The concentration of particles in the drop, on the substrate and in the surrounding medium are given by $a(\vec{x}, t), b(\vec{x}, t)$ and $c(\vec{x}, t)$ (respectively yellow, red and green). Particles may adsorb from the drop onto the substrate or dissolve from the substrate into the surrounding medium with the reaction rates $r_1$ and $r_2$, respectively. Particles that adsorb onto the substrate render it less wettable, thereby causing the drop to move along the wettability gradient with velocity $v$}
\label{fig:system_sketch}
\end{figure}
We now specifically consider drops of a partially wetting solution situated on a flat rigid substrate and enclosed in a vertical gap of height $d$ (Fig.~\ref{fig:system_sketch}). The dissolved molecules may reversibly adsorb from the drop onto the substrate, thereby locally altering its wettability. Away from the drop, molecules may also reversibly desorb from the substrate into a surrounding medium. We assume that both processes can be modeled by reversible monomolecular reactions and formally treat each state of the molecules as a separate chemical species,
\begin{align}
A &\stackrel{r_1}{\leftrightharpoons} B, \label{eq:reaction_AB}\\
B &\stackrel{r_2}{\leftrightharpoons} C, \label{eq:reaction_BC} 
\end{align}
where $A, B$ and $C$ denote the species in the drop, the species on the substrate and the species dissolved in the surrounding medium, respectively. The reaction rates are given by $r_1, r_2 \geq 0$. We do not consider processes like evaporation of the drop and of the chemical on the substrate and therefore have $\tensg{\mathcal{M}}=0$ in Eq.~(\ref{eq:gradient_dynamics_general}). Note that for simplicity, we further assume that the reaction $A\rightleftharpoons C$ does not occur or that its contribution to the dynamics is negligible,~e.g., because of a low reaction rate. In principle, such a reaction could be considered. However, when species $A$ and $C$ are controlled externally via chemostats, this does not influence the remaining dynamics. We assume shallow drops such that spatial gradients in the local film height $h(\vec{x}, t)$ are small,~i.e., we employ a long-wave approximation \citep{BEIM2009rmp,CrMa2009rmp}. The particle number densities for species $A, C$ and $B$ are respectively given by $a(\vec{x}, t), c(\vec{x},  t)$ (particles per unit volume) and $b(\vec{x}, t)$ (particles per unit substrate area). All fields are parametrized in substrate coordinates $\vec{x}=(x,y)^\text{T}$. Note that in this thin-film geometry, $a(\vec{x}, t)$ and $c(\vec{x}, t)$ correspond to vertically averaged quantities \citep{ThTL2013prl, VoTh2024jem}. Furthermore, in this geometry the concentration dynamics in the surrounding medium can only be captured for a \textit{closed} system when there exists a vertical bound,~i.e., drops must be confined to a gap. For details for the analogous case of evaporating drops, see Ref.~\citep{HDJT2023jfm}. The free energy of this system is given by 
\begin{equation}
\begin{split}
\mathcal{F}= &\int\bigg[f(h, b)+hg_a(a)+g_b(b)+(d-h)g_c(c) \\
& +\frac{\gamma}{2}\vert\vec{\nabla}h\vert^2+h\frac{\sigma_a}{2}\vert\vec{\nabla}a\vert^2+\frac{\sigma_b}{2}\vert\vec{\nabla}b\vert^2+(d-h)\frac{\sigma_c}{2}\vert\vec{\nabla}c\vert^2\bigg]\:\text{d}^2 r,
\end{split}\label{eq:free_energy}
\end{equation}
where integration is performed over substrate coordinates. The first contribution $f(h, b) = A_h\left(1+\lambda\frac{b}{b_0}\right)\left(-\frac{1}{2h^2}+\frac{h_a^3}{5h^5}\right)$ represents a mesoscopic wetting energy with the Hamaker constant $A_h>0$ and a single minimum in the film height at $h=h_a$, corresponding to the thickness of the adsorption layer that coexists with the drop in the mesoscopic picture. Note that the thickness of the adsorption layer $h_a$ is independent of $b$. We assume a linear dependence on the adsorbate density $b$ where $\lambda>0$ is a dimensionless proportionality factor and $b_0>0$ is a reference density.\footnote{We do not combine $\lambda$ and $b_0$ into a single factor due to the employed nondimensionalization where $b_0$ is used for scaling, see Appendix \ref{app:nondim}.} This choice implies that the adsorbate renders the substrate \textit{less} wettable. We remark that the corresponding spreading coefficient $S$ must be determined from the consistency of macroscopic and mesoscopic descriptions of thermodynamic equilibrium, in analogy to Ref.~\citep{TSTJ2018l}. This results in an expression for the spreading coefficient that, in general, differs from the one that is obtained for the simple thin-film equation and was given in Refs.~\citep{ThJB2004prl, VoTh2024jem}. We derive an expression for $S$ for the chemostatted passive (relaxational) system in Appendix \ref{app:spreading_coeff}. Here, we have $S<0$ for our choice of $f(h,b)$, corresponding to partial wetting. The next three terms $hg_a, g_b$ and $(d-h)g_c$ represent the free energy contributions of species $A, B$ and $C$, respectively. Note that in contrast to $g_b$ (energy per unit substrate area), the quantities $g_a$ and $g_c$ are energies per unit volume. The remaining terms penalize gradients in the interface height and the densities, where $\gamma>0$ is the surface tension of the liquid and $\sigma_a, \sigma_b, \sigma_c>0$ are interface stiffnesses. When the system is closed,~i.e., the total mass of chemical species $M = m\int [ha + b +(d-h)c]\:\text{d}^2 r$ (with the mass per particle $m$)\footnote{Note that reactions (\ref{eq:reaction_AB}) and (\ref{eq:reaction_BC}) imply that all species have identical masses per particles.} and the liquid volume $V = \int h \:\text{d}^2r$ are conserved and finite, the system is described by a gradient dynamics of the form (\ref{eq:gradient_dynamics_general}) with the dynamical fields $\psi_1=h, \psi_2=ha, \psi_3=b$ and $\psi_4=(d-h)c$, see Refs.~\citep{ThTL2013prl, HDJT2023jfm, VoTh2024jem} for details on the choice of fields. Specifically, the mobility matrix for spatial transport is given by 
\begin{equation}
	\tensg{\mathcal{Q}} = \left(\begin{array}{cccc}
	\frac{h^3}{3\eta} & \frac{h^3a}{3\eta} & 0 & 0 \\ 
	\frac{h^3a}{3\eta} & \frac{h^3a^2}{3\eta}+D_aha & 0 &0 \\
	0 & 0 & D_b b & 0 \\
	0 & 0 & 0 & D_c (d-h) c
	\end{array} \right),\label{eq:mobility}
\end{equation}
where $\eta$ is the dynamic viscosity of the liquid. The upper-left $2\times2$-matrix in (\ref{eq:mobility}) corresponds to a description of thin films of mixtures or suspensions with no-slip and no-penetration boundary conditions at the substrate and with a stress-free ambient medium \citep{ThTL2013prl}. Note that (\ref{eq:mobility}) is positive definite and symmetric. The diffusivity of each species is given by $D_i>0$ with $i=a,b,c$. Specifically for species $B$ and $C$, we assume purely diffusive transport. This follows directly from the underlying boundary conditions of the thin-film problem at the substrate (vanishing liquid velocity) and at the free surface (vanishing stress). The latter boundary condition, corresponding to purely diffusive transport of species $C$, is appropriate when the ambient medium mechanically relaxes much faster than the drop,~e.g., due to large differences in the viscosities of the two fluids. From our choice of fields and Eq.~(\ref{eq:free_energy}), we obtain the variations
\begin{align}
\begin{split}
    \frac{\delta \mathcal{F}}{\delta \psi_1}&=p=\partial_h f -\gamma \vec{\nabla}^2 h+g_a-a \partial_a g_a+ \frac{\sigma_a}{2}\vert \vec{\nabla} a \vert^2 + \sigma_a \frac{a}{h} \vec{\nabla}\cdot (h\vec{\nabla} a), \\
    &\qquad\quad\quad -g_c+c \partial_c g_c- \frac{\sigma_c}{2}\vert\vec{\nabla} c \vert^2 - \sigma_c \frac{c}{(d-h)}\vec{\nabla} \cdot\left[(d-h)\vec{\nabla} c\right],
\end{split}\label{eq:p}\\
    \frac{\delta \mathcal{F}}{\delta \psi_2}&=\mu_a=\partial_a g_a-\frac{\sigma_a}{h}\vec{\nabla}\cdot\left(h\vec{\nabla} a\right),\label{eq:mu_a}\\
    \frac{\delta F}{\delta \psi_3}&=\mu_b=\partial_bg_b+\partial_b f - \sigma_b \vec{\nabla}^2b,\label{eq:mu_b}\\
    \frac{\delta \mathcal{F}}{\delta \psi_4}&=\mu_c=\partial_c g_c-\frac{\sigma_c}{(d-h)}\vec{\nabla}\cdot\left[(d-h)\vec{\nabla} c\right]\label{eq:mu_c},
\end{align}
where $\vec{\nabla}^2$ denotes the two-dimensional Laplace operator. Note that Eq.~(\ref{eq:p}) is the liquid pressure $p$ and Eqs.~(\ref{eq:mu_a})-(\ref{eq:mu_c}) are the chemical potentials $\mu_a, \mu_b$ and $\mu_c$ of $A,B$ and $C$, respectively. The first term in (\ref{eq:p}) corresponds to the disjoining pressure $\Pi(h,b)=-\partial_h f$ and the expressions $g_a-a\partial_ag_a$ and $-g_c+c\partial_cg_c$ are osmotic pressures. The reaction currents in Eq.~(\ref{eq:gradient_dynamics_general}) are now given by $\vecg{\mathcal{R}} = (0, -\mathcal{R}_{ab}, \mathcal{R}_{ab}-\mathcal{R}_{bc}, \mathcal{R}_{bc})^\text{T}$ with
\begin{align}
\mathcal{R}_{ab} &= \bar{r}_1(h) \left[\exp\left(\frac{1}{k_\mathrm{B}T}\frac{\delta \mathcal{F}}{\delta \psi_2}\right)-\exp\left(\frac{1}{k_\mathrm{B}T}\frac{\delta \mathcal{F}}{\delta \psi_3}\right)\right], \label{eq:R_ab}\\
\mathcal{R}_{bc} &= \bar{r}_2(h)  \left[\exp\left(\frac{1}{k_\mathrm{B}T}\frac{\delta \mathcal{F}}{\delta \psi_3}\right)-\exp\left(\frac{1}{k_\mathrm{B}T}\frac{\delta \mathcal{F}}{\delta \psi_4}\right)\right].\label{eq:R_bc}
\end{align}
To account for the circumstance that reactions (\ref{eq:reaction_AB}) and (\ref{eq:reaction_BC}) respectively only take place underneath the drop or away from it, the reaction rates $\bar{r}_1(h)$, $\bar{r}_2(h)$ carry film-height dependencies with $\bar{r}_1(h)=\frac{r_{1}}{2}[1+\tanh(\frac{h-h_0}{\Delta h})]$ and $\bar{r}_2(h)=\frac{r_{2}}{2}[1-\tanh(\frac{h-h_0}{\Delta h})]$. Both rates are smooth step-like functions in $h$ such that reactions (\ref{eq:reaction_AB}) and (\ref{eq:reaction_BC}) are suppressed in the adsorption layer and below the drop, respectively. The parameters $h_0$ and $\Delta h$ give the midpoint and the steepness of the step. Typically, $h_0$ is chosen slightly greater than $h_a$ and $\Delta h$ is small (such that $\bar{r}_1$ is negligible at $h=h_a$). Note that Eqs.~(\ref{eq:R_ab}) and (\ref{eq:R_bc}) nevertheless obey local detailed balance (\ref{eq:local_detailed_balance}). 

The model (\ref{eq:free_energy})-(\ref{eq:R_bc}) represents a closed system that relaxes to thermodynamic equilibrium,~i.e., a resting drop coexisting with an adsorption layer. In this state, assuming that $A$ and $C$ do not phase separate, the densities $a$ and $c$ are uniform while $b$ typically features an interface near the contact line due to the contribution of the wetting energy in (\ref{eq:mu_b}) (cf. Refs.~\citep{ThTL2013prl,VoTh2024jem}). We are now interested in an out-of-equilibrium setting where the particles in the drop are highly abundant and the volume of the surrounding medium is large. The drop and the ambient medium may then be regarded as infinite reservoirs for particles of species $A$ and $C$, respectively,~i.e., they are chemostats with uniform and constant chemical potentials $\mu_a, \mu_c$. For a justification of this approximation for dilute mixtures, see Ref.~\citep{ReEA2025jcp}. Because $A$ and $C$ do not phase separate, their densities are also uniform and constant. Then, the dynamics is described by the reduced system
\begin{align}
\partial_t h &= \vec{\nabla}\cdot\left(\frac{h^3}{3\eta}\vec{\nabla}\frac{\delta F}{\delta h}\right), \label{eq:dt_h_reduced} \\
\partial_t b &= \vec{\nabla}\cdot\left(D_bb\vec{\nabla}\frac{\delta F}{\delta b}\right)+R_{ab}-R_{bc}, \label{eq:dt_b_reduced}
\end{align}
with the reduced free energy
\begin{equation}
F = \int \left[f(h,b)+\frac{\gamma}{2}\vert\vec{\nabla}h\vert^2+g_b(b)+\frac{\sigma_b}{2}\vert\vec{\nabla}b\vert^2\right]\:\text{d}^2 r, \label{eq:F_red}
\end{equation}
and the reaction currents 
\begin{align}
R_{ab} &= \bar{r}_1(h) \left[\exp\left(\frac{\mu_a}{k_\mathrm{B}T}\right)-\exp\left(\frac{1}{k_\mathrm{B}T}\frac{\delta F}{\delta b}\right)\right], \label{eq:R_ab_reduced}\\
R_{bc} &= \bar{r}_2(h)  \left[\exp\left(\frac{1}{k_\mathrm{B}T}\frac{\delta F}{\delta b}\right)-\exp\left(\frac{\mu_c}{k_\mathrm{B}T}\right)\right].\label{eq:R_bc_reduced}
\end{align}
In the following, we employ the energy 
\begin{equation}
g_b(b) = \gamma_{sl}^0+ k_\mathrm{B}T b\left[\ln b/b_0 -1\right], \label{eq:gb}
\end{equation}
where $\gamma_{sl}^0>0$ is the solid-liquid interfacial tension of the bare substrate. We remark that in principle, the constant term  $\gamma_{sl}^0$ may be neglected. Here, we keep it for consistency with the free energy in the macroscopic picture (Appendix \ref{app:spreading_coeff}). Note that our choice (\ref{eq:gb}) does not imply that species $B$ is ideal-gas like due to the interactions with the substrate that are represented by $f(h,b)$. However, Eq.~(\ref{eq:gb}) may be amended to account for,~e.g., close packing of species $B$ on the substrate. In particular, the reaction currents (\ref{eq:R_ab_reduced}), (\ref{eq:R_bc_reduced}) now read explicitly
\begin{align}
R_{ab}&=\bar{r}_1(h) \left(\exp\left[\frac{\mu_a}{k_\mathrm{B}T}\right]-\frac{b}{b_0}\exp\left[\frac{A_h}{k_\mathrm{B}T}\frac{\lambda}{b_0}\left\lbrace-\frac{1}{2h^2}+\frac{h_a^3}{5h^5}\right\rbrace\right]\right), \label{eq:R_ab_explicit}\\
R_{bc}&=\bar{r}_2(h) \left(\frac{b}{b_0}\exp\left[\frac{A_h}{k_\mathrm{B}T}\frac{\lambda}{b_0}\left\lbrace-\frac{1}{2h^2}+\frac{h_a^3}{5h^5}\right\rbrace\right]-\exp\left[\frac{\mu_c}{k_\mathrm{B}T}\right]\right) \label{eq:R_bc_explicit}
\end{align}
Eqs.~(\ref{eq:R_ab_explicit}) and (\ref{eq:R_bc_explicit}) correspond to mass-action type kinetics that are modified by energetic contributions representing the interactions of species $B$ with the substrate. In total, Eqs.~(\ref{eq:dt_h_reduced})-(\ref{eq:dt_b_reduced}) and (\ref{eq:R_ab_explicit}), (\ref{eq:R_bc_explicit}) then differ from the model presented in Ref.~\citep{VoTh2024jem} by the additional reaction term $-R_{bc}$ that represents the exchange of adsorbate on the substrate with the ambient medium. The corresponding system is open but may nevertheless be passive,~i.e., it may relax to a local minimum of an appropriate thermodynamic potential. Here, this is the case for $\mu_a=\mu_c=\mu_0$, where the semi-grand free energy\footnote{This definition is analogous to the introduction of the standard grand potential $\Omega = F-\sum_\alpha\int\mu_\alpha n_\alpha\:\text{d}^2r$ for an open system without chemical reactions, where $n_\alpha$ and $\mu_\alpha$ are the particle number density and chemical potential of species $\alpha$, respectively. However, when chemical reactions occur, already for a closed system not all particle numbers are control parameters of the equilibrium system. Instead, the equilibrium state is controlled only via the set of chemically conserved quantities that are encoded in the stoichiometry of the reactions \citep{AAFE2024jcp}. Specifically, the coefficients $c_\alpha$ of any conserved quantity $\int \sum_{\alpha=1}^N c_\alpha n_\alpha\:\text{d}^2x$ must fulfill the condition $\sum_{\alpha=1}^N(\nu^+_{r\alpha}-\nu^-_{r\alpha}) c_\alpha=0$ for all  $r=1,\ldots, R$. Then, when one transitions to an open equilibrium system via the exchange of matter with external chemostats, to obtain the proper thermodynamic potential one must only consider the formerly conserved quantities that are exchanged with the environment. For the here considered system, there is only one conserved quantity that is exchanged with the chemostats, namely the total particle number $N_T=\int ha+b+(d-h)c\:\text{d}^2 r$ (i.e., we have $c_\alpha=1$ for all $\alpha$). Equivalently, one may also use the total mass of chemical species $M$. Further, in the present case chemical equilibrium implies $\mu_a= \mu_c=\mu_b=\mu_0$ and one obtains in analogy to the construction of the grand potential instead the semi-grand free energy $F_s = F-\mu_0 N_T$. This definition can be extended to the nonequilibrium case, where we instead write $F_s=F-\mu_0\int b\:\text{d}^2 r$ since $a$ and $c$ are assumed constant due to chemostatting and the liquid volume is conserved. For details on the general thermodynamic framework, we refer to \citep{AAFE2024jcp}.} $F_s=F-\mu_0\int b\:\text{d}^2 r$ is minimized (cf. Ref.~\cite{AAFE2024jcp}). Otherwise, the system is kept permanently out of equilibrium. In particular, when $\mu_a>\mu_c$ particles tend to adsorb onto the substrate when the drop is present and dissolve into the ambient medium away from the drop. That is, the drop and the surrounding fluid respectively represent a reactive source (fuel reservoir) and sink (waste reservoir) for particles on the substrate. For sufficiently strong nonequilibrium driving, drops may then self-propel across the substrate with some velocity $v$, driven by a localized wettability gradient across the drop that is sustained by reactions (\ref{eq:reaction_AB}) and (\ref{eq:reaction_BC}). We give a nondimensionalization of Eqs.~(\ref{eq:dt_h_reduced})-(\ref{eq:R_bc_reduced}) in Appendix \ref{app:nondim}. In the following, all quantities are nondimensional.
\begin{figure}[H]
    \centering
     \includegraphics[scale=0.5]{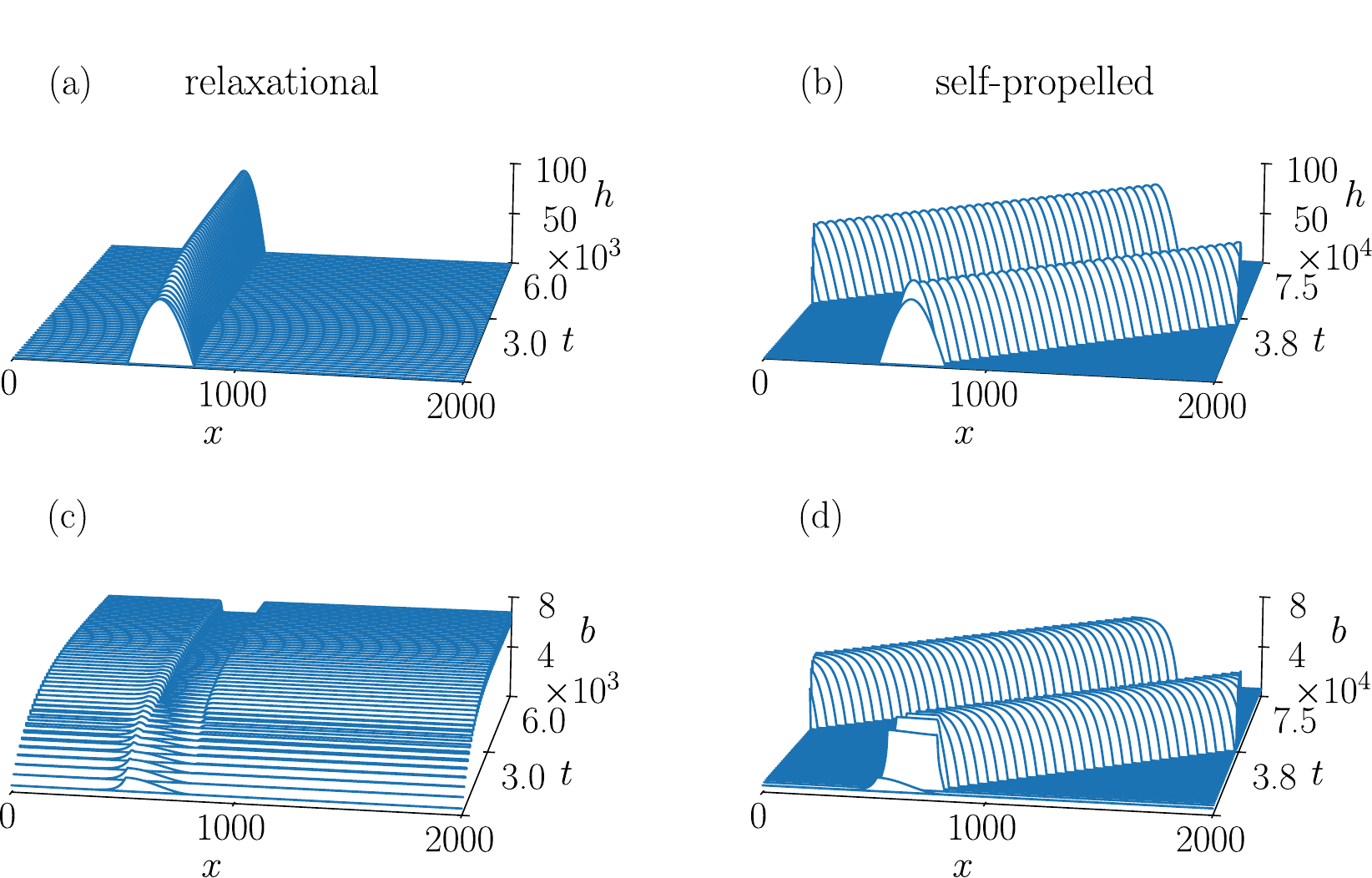}
    \caption{Comparison of (a,~c) relaxational dynamics converging to a resting drop and (b,~d) persistent out-of-equilibrium dynamics converging to a self-propelled drop. Panels (a) and (b) [(c) and (d)] show space-time representations of the drop profile [adsorbate profile] for an initially symmetric drop with a small adsorbate concentration gradient (i.e., a wettability gradient) underneath the drop. In the passive relaxational case, the initial concentration gradient is equilibrated and the overall adsorbate concentration gradually increases. In the contact line regions, interfaces are formed in the concentration profile as the system relaxes to thermodynamic equilibrium, where the drop is at rest. In the persistently out-of-equilibrium case, the initial concentration gradient beneath the drop is amplified while adsorbate is continuously removed from the surrounding substrate. Ultimately, drops self-propel across the substrate with constant velocity. The external chemical potentials are (a,~c) $\mu_a=\mu_c=\ln 6$ and (b,~d) $\mu_a=\ln 6, \mu_c=\ln 0.5$. The remaining parameters are $W=1, D_b=0.001, r_1=0.001, r_2=0.001, h_0 = 2.4, \Delta h = 0.6,\sigma_b= 1, \lambda = 0.5$. The periodic computational domain is [0, 2000]}
    \label{fig:passive_active}
\end{figure}
\begin{figure}[H]
    \centering
     \includegraphics[scale=0.6]{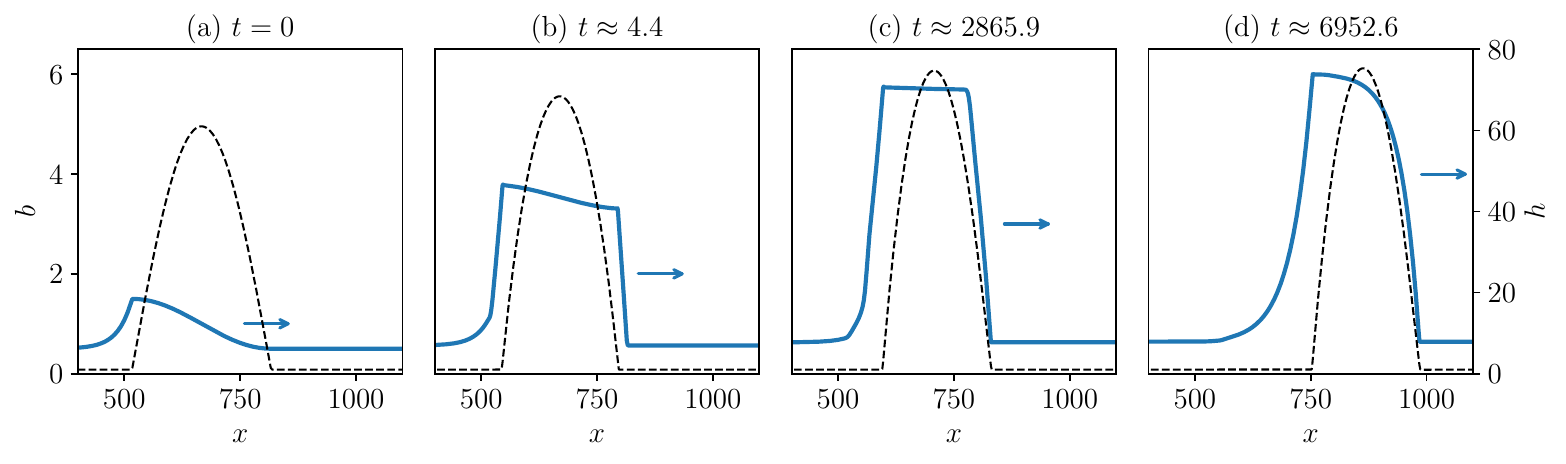}
    \caption{Snapshots of the adsorbate profile (blue solid lines) and the drop profile (black dashed lines) at different times in the persistently out-of-equilibrium case in Fig.~\ref{fig:passive_active}. Panel (a) corresponds to the initial condition. As time progresses, the initially small concentration gradient beneath the drop is amplified [panels (b) and (c)] and develops into a profile that increases across the drop and then decays behind it [panel (d)]. In (d), the profile corresponds to stable uniform drop motion}
    \label{fig:b_profile_evolution}
\end{figure}
\subsection{Numerical Results}
We numerically investigate the model (\ref{eq:dt_h_reduced})-(\ref{eq:R_bc_reduced}) via time simulations employing finite-element open source library oomph-lib \cite{HeHa2006} for a one-dimensional periodic domain [0,2000] that is discretized using a spatially adaptive mesh with linear Lagrange elements. For temporal discretization, we use a second-order backward differentiation scheme [BDF(2)] with adaptive time stepping.

 Fig.~\ref{fig:passive_active} shows a comparison of the passive relaxational case ($\mu_a = \mu_c$) converging to thermodynamic equilibrium,~i.e., a resting drop,  and persistently out-of-equilibrium dynamics converging to a self-propelled drop ($\mu_a>\mu_c$). Both simulations are initiated with symmetric drop profiles and weakly asymmetric profile in $b$,~i.e., with a small initial wettability gradient underneath the drop [Fig.~\subref{fig:b_profile_evolution}{(a)}]. In the passive case [Figs.~\subref{fig:passive_active}{(a)} and \subref{fig:passive_active}{(c)}], interfaces in $b$ form in the contact line regions and drops approximately remain at rest as the system relaxes to thermodynamic equilibrium. Note that the interfaces in $b$ appear due to the wetting energy contribution in the chemical potential (\ref{eq:mu_b}). This is analogous to the passive case in Ref.~\citep{VoTh2024jem}. In the persistently out-of-equilibrium case [Figs.~\subref{fig:passive_active}{(b)} and \subref{fig:passive_active}{(d)}], the initial wettability gradient is gradually amplified (Fig.~\ref{fig:b_profile_evolution}) as adsorbate accumulates beneath the drop while it is continuously removed from the surrounding substrate. Ultimately, drops then persistently self-propel across the substrate with constant velocity.
\begin{figure}[H]
    \centering
     \includegraphics[scale=0.5]{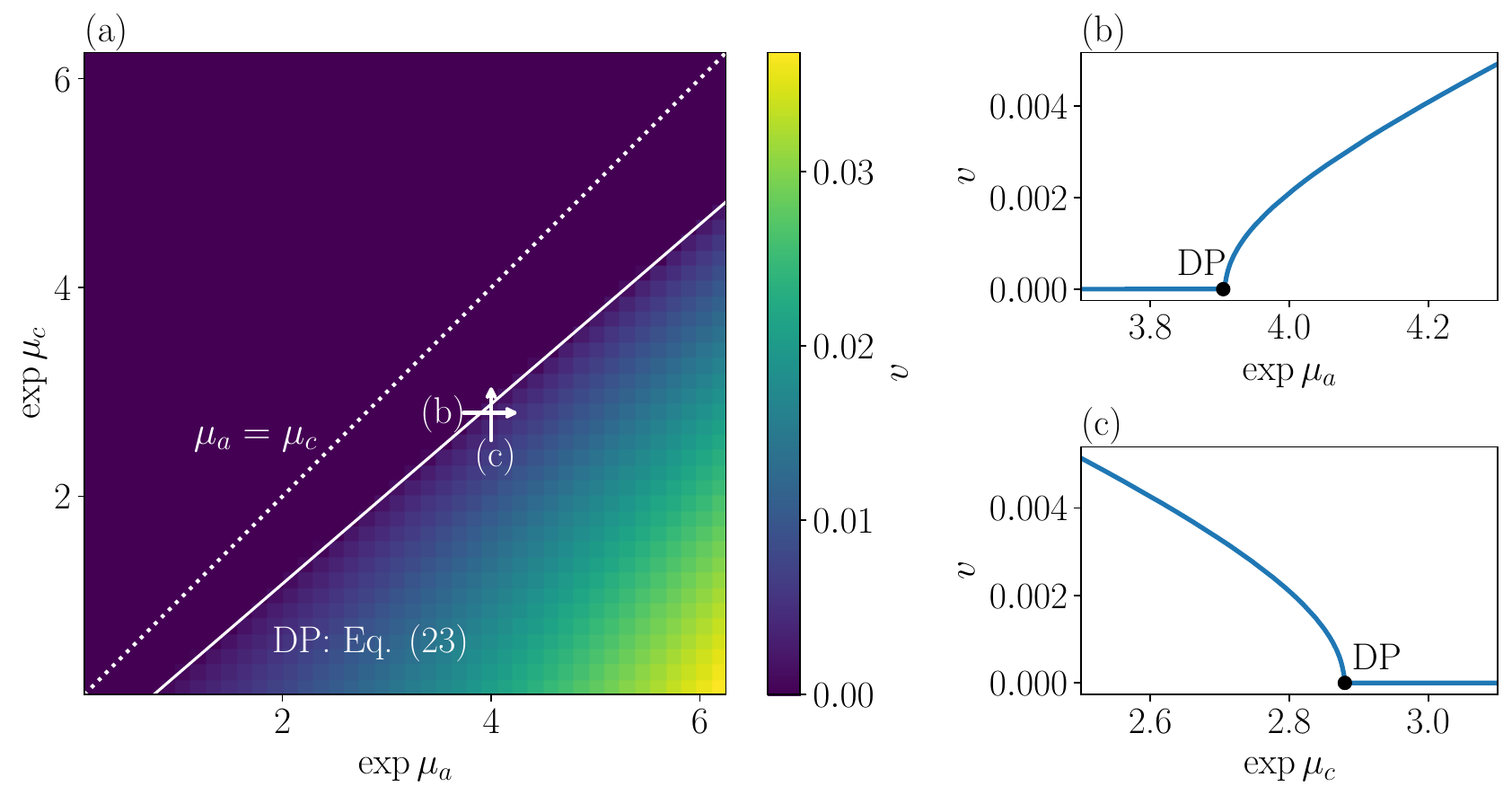}
    \caption{Emergence of drop self-propulsion in dependence of the external chemical potentials $\mu_a$ and $\mu_c$ that characterize the chemostats. (a) Parameter scan of time simulations in the $(e^{\mu_a}, e^{\mu_c})$-plane. Colors indicate the drop velocity in the converged state. The passive case $\mu_a=\mu_b$ is represented as a white dotted line. Self-propelled drops emerge at a line of supercritical drift-pitchfork bifurcations (DP). An estimate of the loci of the DPs is given by Eq.~(\ref{eq:DP_onset_analytical}) [white solid line]. Panels (b) and (c) respectively show the velocity $v$ in dependence of $e^{\mu_a}$ and $e^{\mu_c}$ near the bifurcation,~i.e., for horizontal and vertical cuts in (a) (visualized as white arrows). The drift-pitchfork bifurcations are marked with black filled circles. The remaining parameters and the initial condition are as in Fig.~\ref{fig:passive_active}} 
    \label{fig:mu_parameter_scan}
\end{figure}
It is possible to transition between cases of convergence to a resting and to a self-propelled drop by varying the chemostats,~i.e., either $\mu_a$ or $\mu_c$ [Fig.~\subref{fig:mu_parameter_scan}{(a)}]. Specifically, self-propulsion can occur for $\mu_a>\mu_c$ when crossing a threshold in the $(\mu_a, \mu_c)$-plane (that does not coincide with $\mu_a = \mu_c$). In this parameter plane the threshold corresponds to a line of supercritical drift-pitchfork bifurcations (DP) \citep{GGGC1991pra,FaDT1991jpi}, where broken-parity self-propelled drops emerge [Figs.~\subref{fig:mu_parameter_scan}{(b)} and \subref{fig:mu_parameter_scan}{(c)}]. Notably, the velocity vanishes at the DP and increases as $v\propto \vert e^{\mu_i}-e^{\mu_i^0}\vert^{1/2}$ beyond the instability threshold where $i=a, b$ and $\mu_a^0, \mu_b^0$ are the parameter values at the bifurcation.
In our simulations, we find that the onset of drop motion corresponds to a critical wettability difference between the drop and the surrounding adsorption layer, which corresponds to the concentration difference $\Delta b^0 = b_d^0-b_a^0$. Here, $b_d^0, b_a^0$ are the concentrations beneath the drop and in the adsorption layer at onset, respectively. Note that $\Delta b^0>0$. Physically, this reflects that drop motion only occurs for finite wettability gradients across the drop. Note that this is not the case for drops subjected to externally imposed wettability gradients, where drop motion already occurs for arbitrarily small gradients. Conceptually, the difference between this case and the present scenario is that here the wettability profile is dynamic and governed by the time evolution of $b$ (which in turn couples to the dynamics of the drop). This is further reflected in the circumstance that in the steady state, there are no gradients in the liquid pressure $p$ and therefore only the dynamics of $b$ contributes to the total dissipation.\footnote{This can be seen from the steady state condition for the film height profile $j_0=-\frac{h^3}{3}\partial_x p$, where $j_0$ is the uniform and constant flux in the steady state. Integrating over the domain of length $L$, one obtains $\int_0^L \frac{3 j_0}{h^3}=p(x=0)-p(x=L)=0$, where the second equality follows from periodic boundaries. Because we have $h>0$ everywhere, it follows that $j_0=0$ and therefore $\partial_x p =0$.} In our simulations, we find that this steady state dissipation is localized in the contact line region since $b$ is usually uniform beneath the drop bulk and in the adsorption layer, such that diffusive and reactive fluxes vanish separately in the respective regions (i.e., there they do not contribute to dissipation). We expect that the corresponding nonequilibrium fluxes in the contact line region are directly connected to the observation that the onset of drop motion occurs at a finite wettability contrast. However, an in-depth discussion of the underlying physical mechanism requires a detailed understanding of the dynamics in the contact line region and is likely highly nontrivial. Nevertheless, an estimate for the location of the onset of motion in the $(e^{\mu_a}, e^{\mu_c})$-plane can be obtained as follows. In the steady state, for sufficiently fast reactions the concentration profile of $b$ is uniform away from the contact line region and the corresponding chemical potential in each region is simply determined by the values imposed by the chemostats. In particular at bifurcation, we then have $\mu_{b , d}^0= \mu_a^0$ and $\mu_{b, a}^0= \mu_c^0$, where $\mu_{b, d}^0,\mu_{b, a}^0$ are the respective chemical potentials of $B$ beneath the drop and in the adsorption layer at bifurcation. Then, using Eqs.~(\ref{eq:mu_b}) and (\ref{eq:gb}), we have at onset
\begin{equation}
e^{\mu_a^0}-e^{\mu_c^0-\partial_b f\eval_{h=h_a}}=\Delta b^0,
\end{equation}
or equivalently,
\begin{equation}
e^{\mu_c^0} = e^{\partial_b f\eval_{h=h_a}}e^{\mu_a^0}-e^{\partial_b f\eval_{h=h_a}}\Delta b^0, \label{eq:DP_onset_analytical}
\end{equation}
with $\partial_b f \eval_{h=h_a}= -W\lambda\frac{3}{10}<0$ in nondimensional units (see Appendix \ref{app:nondim}). Equation (\ref{eq:DP_onset_analytical}) corresponds to a straight line in the $(e^{\mu_a}, e^{\mu_c})$-plane with the slope $e^{\partial_b f\eval_{h=h_a}}<1$ and the root $e^{\mu_a^0}=\Delta b^0$. We determine the critical contrast $\Delta b^0>0$ directly from numerical simulations. As shown in Fig.~\subref{fig:mu_parameter_scan}{(a)}, the onset of motion is well-captured by (\ref{eq:DP_onset_analytical}) and is therefore determined by $\partial_b f \eval_{h=h_a}$ and $\Delta b^0$, which reflect the interactions of species $B$ with the substrate and the required wettability contrast for drop motion, respectively. Note that these quantities are not necessarily independent,~i.e., we expect that $\Delta b^0$ depends,~e.g., on the substrate interactions and the drop volume. Further, we remark that the above calculation [more generally, the analysis presented in Fig.~\subref{fig:mu_parameter_scan}{(a)}] requires that nonequilibrium driving is captured via chemostats in a thermodynamically consistent manner and is therefore not possible for previous models of chemically driven running drops \citep{ThJB2004prl, JoBT2005epje}.

\begin{figure}[H]
    \centering
     \includegraphics[scale=0.7]{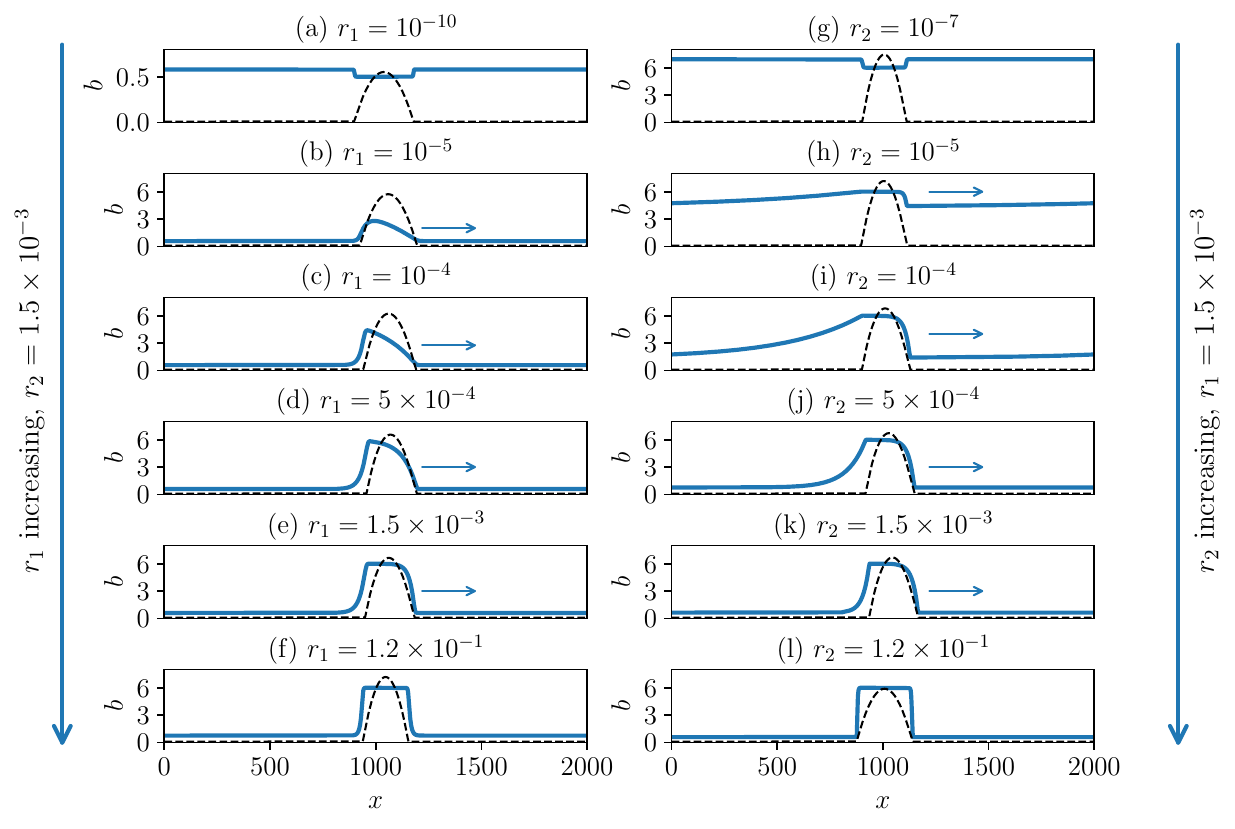}
    \caption{Concentration profiles (blue solid lines) of stationary [(a), (g), (f), (l)] and self-propelled drops [(b)-(e), (h)-(k)] in dependence of the reaction rates $r_1$ (left) and $r_2$ (right). The corresponding drop profiles are additionally superimposed (black dashed lines). Note that for the drop profile identical scales are used in all panels. Arrows indicate moving drops. In the left [right] column, the reaction rate $r_1$ [$r_2$] increases from top to bottom. As $r_1$ is increased at fixed $r_2=1.5\times 10^{-3}$, the overall concentration beneath the drop is increased until a saturation value is reached. This corresponds to a plateau in $b$ across the drop, see (e). When $r_2$ is increased at fixed $r_1=1.5\times 10^{-3}$, the adsorbate concentration on the substrate away from the drop is reduced, reaching a lower plateau value, see (j) and (k). When values of either $r_1$ or $r_2$ are too small [(a), (g)] or too large [(f), (l)], drops are stationary.  The remaining parameters are as in the self-propelled case in Fig.~\ref{fig:passive_active}, except for $D_b=0.1$ and $\Delta h=0.35$ in (a)-(f)}
    \label{fig:r1_r2_profiles_overview}
\end{figure}

In addition to the dependence on the chemostats, the drop and concentration profiles are also affected by the speed of the chemical reactions. We show in Fig.~\ref{fig:r1_r2_profiles_overview} how the rates $r_1$ and $r_2$ of reactions (\ref{eq:reaction_AB}) and (\ref{eq:reaction_BC}) influence the adsorbate profiles on the substrate for the converged self-propelled drops. Specifically, an increase in $r_1$ corresponds to a faster adsorption of particles from the drop at the substrate. This results in an overall increase of adsorbate density beneath the drop [Fig.~\subref{fig:r1_r2_profiles_overview}{(a)-(f)}]. For fast adsorption, the concentration underneath the drop saturates to a plateau value given by $R_{ab}=0$ [Figs.~\subref{fig:r1_r2_profiles_overview}{(e)},~\subref{fig:r1_r2_profiles_overview}{(f)}]. However, when the adsorbate deposition is too slow or too fast [Figs.~\subref{fig:r1_r2_profiles_overview}{(a)},~\subref{fig:r1_r2_profiles_overview}{(f)}], the adsorbate profile becomes symmetric and drops remain at rest. Similarly, a change in $r_2$ increases the rate of dissolution into the surrounding medium and, correspondingly, the adsorbate concentration away from the drop decreases with increasing $r_2$. For sufficiently large values of $r_2$, the adsorbate profile away from the drop also saturates to a lower plateau value determined by $R_{bc}=0$ [Figs.~\subref{fig:r1_r2_profiles_overview}{(j)-(l)}].  As before, for very slow or very fast dissolution drops do not self-propel [Figs.~\subref{fig:r1_r2_profiles_overview}{(g)} and \subref{fig:r1_r2_profiles_overview}{(l)}]. Note that the former case ($r_2\rightarrow 0$) recovers the model presented in Ref.~\citep{VoTh2024jem},~i.e., the case of an irreversible substrate coverage. 

\begin{figure}[H]
    \centering
     \includegraphics[scale=0.35]{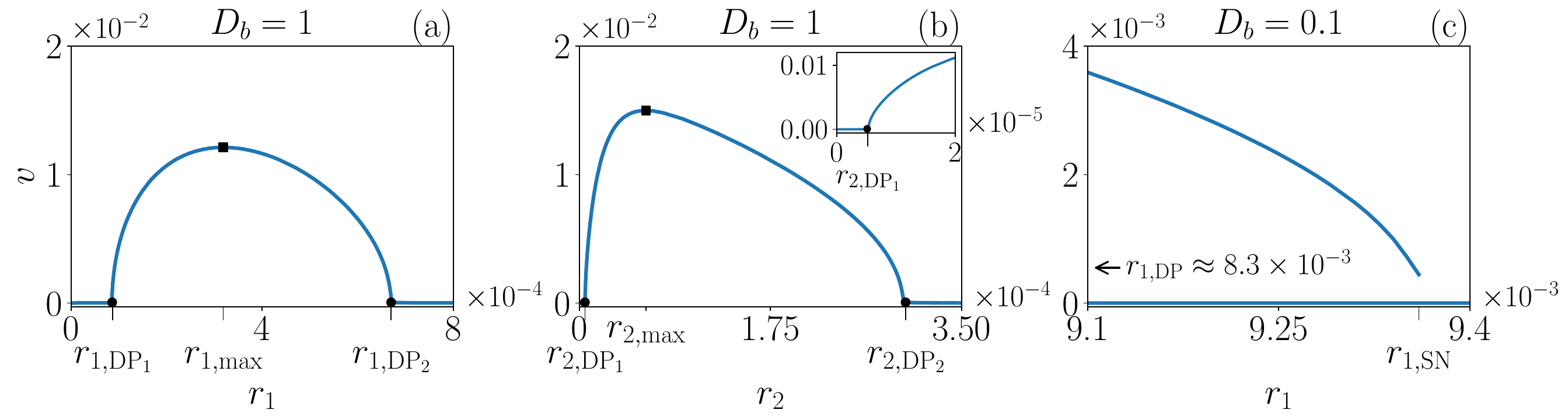}
    \caption{Drop velocity $v$ in dependence of $r_1$ and $r_2$ for different adsorbate diffusivities $D_b$. Panels (a) and (b) respectively show the dependence on $r_1$ and $r_2$ for $D_b=1$. Panel (c) shows the dependence on $r_1$ for $D_b=0.1$. In (a) [(b)], supercritical drift-pitchfork bifurcations (black circles) occur at $r_{1, \mathrm{DP}_1}$ and $r_{1, \mathrm{DP}_2}$ [$r_{2, \mathrm{DP}_1}$ and $r_{2, \mathrm{DP}_2}$]. Furthermore, the velocity is respectively maximal at $r_{1, \mathrm{max}}$ and $r_{2, \mathrm{max}}$ (black squares). The inset in (b) shows a magnification of the left drift-pitchfork. In (c), the bifurcation is subcritical, as indicated by the multistability of stationary ($v=0$) and self-propelled drops ($v\neq 0$). We expect that a saddle-node bifurcation of the self-propelled drop branch occurs at $r_{1, \mathrm{SN}}$, where it turns towards lower $r_1$, ultimately reconnecting to the branch of resting drops at a subcritical drift-pitchfork at $r_{1,\mathrm{DP}}\approx 8.3\times 10^{-2}$, that is, far left
of the shown range. The reaction rates are (a) $r_2=1.6\times 10^{-4}$, (b) $r_1=2.6\times 10^{-4}$ and (c) $r_2=1.5\times 10^{-3}$. In (c), we use $\Delta h=0.35$. The remaining parameters are as in the self-propelled case in Fig.~\ref{fig:passive_active}}
    \label{fig:r_bifurcation_diagrams}
\end{figure}

As drops only self-propel when reactions (\ref{eq:reaction_AB}) and (\ref{eq:reaction_BC}) are not too fast or too slow, we also expect DP when using $r_1$ and $r_2$ as control parameters. Indeed, we find corresponding transitions between resting and self-propelled drops [Figs.~\subref{fig:r_bifurcation_diagrams}{(a)-(b)}]. Varying either $r_1$ or $r_2$, drop self-propulsion occurs in between two critical values of the reaction rates, each corresponding to a supercritical DP. From Figs.~\subref{fig:r_bifurcation_diagrams}{(a)-(b)}, we further see that the self-propulsion speed is maximal for some intermediate values of $r_1$ and $r_2$, respectively given by $r_{1, \mathrm{max}}$ and $r_{2, \mathrm{max}}$. Finally, while in Fig.~\subref{fig:r_bifurcation_diagrams}{(a)-(b)} all DP are supercritical, we note that changes in the adsorbate diffusivity $D_b$ can result in a subcritical DP, as indicated by the multistability of stationary and self-propelled drops in Fig.~\subref{fig:r_bifurcation_diagrams}{(c)}. We expect that the branch of moving drops undergoes a saddle-node bifurcation at $r_{1,\mathrm{SN}}$ [Fig.~\subref{fig:r_bifurcation_diagrams}{(c)}], where it turns toward lower values of $r_1$, and ultimately connects to a subcritical DP at $r_{1,\text{DP}}$ far left of the right limit of  Fig.~\subref{fig:r_bifurcation_diagrams}{(c)}. Similar behavior when varying the adsorbate diffusivity has previously been reported for models without full thermodynamic consistency and with reversible substrate coverage \citep{JoBT2005epje}. 

\section{Conclusion}\label{sec:conclusion}
In conclusion, we have presented a thermodynamically consistent model of chemically driven running drops for the case of reversible substrate coverage. In our derivation of the model, we have started with a full gradient dynamics description of a closed system by considering sessile drops in a surrounding fluid medium confined to a vertical gap. This has allowed us to capture the dynamics of the drop and of the different chemical species in the drop, at the substrate and in the ambient fluid. By then chemostatting the species in the drop and in the ambient fluid, we have transitioned to a description of only the dynamics of the drop profile and the adsorbate profile at the substrate. The thus obtained model represents an extension of the one presented for the irreversible case in Ref.~\citep{VoTh2024jem}, with additional exchange between the substrate and the ambient medium. In contrast to Ref.~\citep{VoTh2024jem}, the present model admits genuine sustained nonequilibrium when the external chemical potentials are distinct, thereby allowing for persistent drop self-propulsion,~e.g., on a periodic domain, driven by self-sustained local wettability gradients. 

We have studied the emerging drop self-propulsion using direct numerical simulations. Notably, the employed description involving chemostats allows for the study of drop self-propulsion in dependence of the external chemical potentials,~i.e., a well-defined nonequilibrium driving. When the chemostatted species are ideal,~i.e., purely entropic, the external chemical potentials may be directly translated into concentrations, representing a common experimental control parameter \citep{DoOn1995prl,LeKL2002pre}. Within previous models \citep{ThJB2004prl, JoBT2005epje}, these (implicitly) externally imposed concentrations are inaccessible. We find that by varying the chemostats, self-propelled drops emerge via (supercritical) drift-pitchfork bifurcations. Physically, this bifurcation is related to a critical finite wettability difference between the substrate regions underneath the drop and away from it. This is in contrast to the case of externally imposed wettability gradients, where arbitrarily small gradients already drive drop motion. Further, drift-pitchfork bifurcations also occur when the reaction rates are employed as control parameters. Specifically, drops are at rest when either reaction proceeds too fast or too slow. Here, depending on the speed of adsorbate diffusion, the bifurcation may be super- or subcritical.

The present study may be extended in several directions. First, one may employ numerical continuation methods \citep{KrauskopfOsingaGalan-Vioque2007,DTZF2017n,SaNe2016epjt,DWCD2014ccp,DoKK1991ijbc} to further investigate the underlying bifurcation structure, thereby also obtaining branch sections of unstable states that are inaccessible via direct numerical simulations. Here, it is of particular interest how the underlying multi-parameter bifurcation structure differs between the present thermodynamically consistent approach and previous models,~e.g., via the signatures of higher-codimension points or bifurcations. Furthermore, the present model is readily extended to capture more involved systems. For instance, in our description we have neglected the influence of advection in the surrounding fluid assuming a large difference in viscosity. For a description of the case where both fluids (drop, ambient medium) are of comparable viscosity, the present model could be extended in analogy to Ref.~\citep{MPBT2005pf}, resulting in an amended mobility matrix $\tensg{\mathcal{Q}}$. Similarly, one could include experimentally common liquid substrates \citep{NSKY2005pre,KuPr2007aci,SYM2023sr,SSK2017sm} along the lines of \citep{DiTh2025prf}, where one then also needs to consider the solutal Marangoni effect \citep{ThAP2012pf, ThAP2016prf}. The interplay of thereby caused Marangoni flows with chemical reactions of different surfactant species can result in intricate modes of drop motion when kept permanently out of equilibrum via chemostats \cite{VoTh2025prf,VoTh2025arxiv}. Furthermore, while in the present work we have employed film height-dependent reaction rates to model reactions that only occur beneath or away from the drop, comparable effects can be achieved by amending the free energy and assuming constant reaction rates, instead. The resulting reactive currents are then of similar form as the ones introduced here, but the corresponding additional pressure and chemical potential contributions influence both advective and diffusive transport. How such an approach influences the underlying physics of,~e.g., drop self-propulsion is an open question.
Additionally, more involved reaction schemes with higher-order reactions may be explored, representing,~e.g., active cooperative binding to the substrate similar to effects in cell adhesion \citep{KRLW2009sm, WAKR2009sm, BlGo2021prx}. Depending on the complexity of the binding process, one may expect sustained dynamic wettability patterns beneath the drop, possibly leading to highly complex forms of drop motion. Beyond the given examples of reactive wetting, also other systems and models of reactive hydrodynamics should be reconsidered with a focus on thermodynamic consistency of the chemo-mechanical coupling and controlled out-of-equilibrium settings. Examples include the hydrodynamics of droplets that are immersed in an ambient liquid and undergo a Belousov-Zhabotinsky reaction \cite{KYNS2011pre,ChSD2022pre}, self-oscillating gels and vesicles \cite{KTAY2017mh}, and the interplay of chemical reactions and bulk flow driven by emerging gradients of mass density \cite{EcGr1999prl,BuDe2017c}.

\section*{Competing Interests}
The authors have no competing interests to declare.

\section*{Author Contributions}
Conceptualization: FV, UT; Methodology: JN, FV, UT; Formal Analysis: JN, FV; Investigation: JN; Interpretation: JN, FV, UT; Software: JN, FV; Visualization: JN; Writing - original draft: FV; Writing - review and editing: JN, FV, UT; Supervision: FV, UT; Funding Acquisition: UT, Project Administration: UT;

JN and FV contributed equally to this work.

\section*{Acknowledgements}
Part of the calculations for this publication were performed on the HPC cluster PALMA II
of the University of Münster, subsidized by the Deutsche Forschungsgemeinschaft (DFG) (INST
211/667-1). We acknowledge financial support by the DFG via the grant no.\ TH 781/12-2 within
SPP 2171. UT would like to thank the Kavli Institute for Theoretical Physics (KITP), Santa Barbara, for support and hospitality during the programme \textit{Active Solids} where part of the work was undertaken. This research was supported in part by grant NSF PHY-2309135 to KITP. 

\section*{Data Availability}
The data that support the findings of this study as well as the python codes for creating the figures are publicly available at \citep{NiVT2026zn}. 

\appendix
\section{Spreading coefficient}\label{app:spreading_coeff}
Here, we determine the spreading coefficient of the chemostatted passive system given by Eqs.~(\ref{eq:dt_h_reduced})-(\ref{eq:R_bc_reduced}) with $\mu_a=\mu_b=\mu_0$. To this end, we follow the approach presented in Ref.~\citep{TSTJ2018l}. Essentially, we compute the conditions for thermodynamic equilibrium (corresponding to a symmetric drop) in both the macroscopic and the mesoscopic picture, resulting in two formulations of Young's law. A comparison then yields an expression for the spreading coefficient in terms of mesoscopic quantities. In the following, we again restrict ourselves to one-dimensional substrates. First, we consider the macroscopic description based on the the semi-grand free energy
\begin{equation}
F_s^\text{macro} = \int_0^{R_c} \:\text{d}x\left[\gamma \xi +g_{sl}(b)-ph-\mu_0b\right]+\int_{R_c}^\infty \:\text{d}x  \left[g_{sg}(b)-\mu_0b\right]+\lambda_h h(R_c), \label{app:eq:F_macro}
\end{equation}
where $\xi=(1+(\partial_xh)^2)^{1/2}$ is the metric factor for the drop surface. Macroscopically, the three-phase contact line has a sharp position given by $x=R_c$. For symmetry reasons, we consider only the half-domain $[0, \infty)$. Equation (\ref{app:eq:F_macro}) is the macroscopic full-curvature equivalent of the semi-grand free energy in the mesoscopic picture that is minimized for $\mu_a=\mu_c=\mu_0$. In (\ref{app:eq:F_macro}), the first term $\gamma\xi$ represents the energetic contribution of the liquid-gas interface. Note that by transitioning from the full-curvature formulation to the long-wave limit $\vert \partial_x h\vert\ll1$, this contribution becomes $\frac{\gamma}{2}\left(\partial_x h\right)^2$ (dropping the constant contribution $\gamma$). The terms $g_{sl}(b)$ and $g_{sg}(b)$ are the energetic contributions of the solid-liquid and the solid-gas interface, respectively. They are, in general, functions of the particle number density on the substrate. The Lagrange multipliers $p$ and $\lambda_h$ correspond to liquid volume conservation and to an imposed vanishing film height at the contact line, respectively. Specifically, $p$ is the liquid pressure. Note that $\mu_0$ is the externally imposed chemical potential of the chemostats and is \textit{not} a Lagrange multiplier. The variation of $F_s^\text{macro}$  is
\begin{equation}
\begin{split}
\delta F_s^\text{macro} =&\int_0^{R_c}\:\text{d}x\left[-\gamma\kappa-p\right]\delta h(x)+\int_0^{R_c}\:\text{d}x\left[\partial_bg_{sl}-\mu_0\right]\delta b(x)+\int_{R_c}^\infty\:\text{d}x\left[\partial_bg_{sg}-\mu_0\right]\delta b(x) \\
& +\left[\gamma\frac{\partial_x h}{\xi}+\lambda_h\right]\delta h({R_c}) + \left[\gamma \xi -ph+g_{sl}-\mu_0b_d-(g_{sg}-\mu_0 b_a)+\lambda_h\partial_xh\right]\delta R_c, 
\end{split} \label{app:eq:macro_variation}
\end{equation}
where we have introduced the curvature $\kappa = \partial_{xx}h/\xi^3$. Note that the final two terms in (\ref{app:eq:macro_variation}) are evaluated in the limit $x\rightarrow R_c$. In particular, the concentrations $b_d$ and $b_a$ correspond to the evaluation of $b$ in the limits $x\nearrow R_c$ and $x \searrow R_c$, respectively. Importantly, they need not be identical,~i.e., there may be a concentration discontinuity at the contact line in the macroscopic picture. At thermodynamic equilibrium we have $\delta F^\text{macro}_s=0$, which implies the relations 
\begin{align}
p& = -\gamma\kappa, \hspace{0.25cm} x\in\left[0, R_c\right), \label{app:eq:laplace} \\
\mu_0 &= \partial_b g_{sl}, \hspace{0.25cm} x\in\left[0, R_c\right), \label{app:eq:mu0_1}\\
\mu_0 &= \partial_b g_{sg}, \hspace{0.25cm} x\in\left[R_c, \infty\right), \label{app:eq:mu0_2} \\
\lambda_h &= -\gamma\frac{\partial_x h}{\xi}, \hspace{0.25cm}\text{at}\hspace{0.25cm} x=R_c, \label{eq:app:eq:lambda_h}\\
\gamma\xi &= ph-(g_{sl}-\mu_0b_d)+g_{sg}-\mu_0 b_a-\lambda_h\partial_x h, \hspace{0.25cm}\text{at}\hspace{0.25cm} x=R_c. \label{app:eq:var_R}
\end{align}
Equation (\ref{app:eq:laplace}) is the Laplace law. From Eqs.~(\ref{app:eq:mu0_1}) and (\ref{app:eq:mu0_2}), we see that $b$ is spatially uniform underneath the drop and away from it, assuming that $B$ does not phase separate (i.e., $\partial_bg_{sl}$ and $\partial_bg_{sg}$ are invertible). Then, the corresponding densities below and away from the drop are given by $b_d$ and $b_a$, respectively. From Eqs.~(\ref{app:eq:var_R}) and (\ref{eq:app:eq:lambda_h}), using the constraint $h(R_c)=0$ and the relation $1/\xi=\cos\theta_e$ at $x=R_c$ with the equilibrium contact angle $\theta_e$, we obtain the macroscopic Young law 
\begin{equation}
\gamma(\cos\theta_e-1) = \gamma_{sg}-\gamma_{sl} -\gamma=S,\label{app:eq:young_macro}
\end{equation}
where we have identified the equilibrium interfacial tensions,
\begin{align}
\gamma_{sl} & =g_{sl}(b_d)-\mu_0b_d,\label{app:eq:gamma_sl_macro} \\
\gamma_{sg} &= g_{sg}(b_a)-\mu_0b_a, \label{app:eq:gamma_sg_macro}
\end{align}  
as semi-grand free energy densities (analogous to the grand potential density in Ref.~\citep{TSTJ2018l}). The spreading coefficient is denoted by $S$.

Next, we turn to the mesoscopic picture where wettability is encoded in the wetting energy and the contact line becomes a contact line region. There, the semi-grand free energy is given by
\begin{equation}
F_s^\text{meso}= \int_0^\infty\:\text{d}x\left[f(h,b)+g_b(b)+\gamma\xi+\frac{\sigma_b}{2}(\partial_x b)^2-\mu_0b-ph\right] \label{app:eq:F_meso}
\end{equation}
Again, $p$ is the Laplace multiplier representing liquid volume conservation and $\mu_0$ is the externally imposed chemical potential of the chemostats. The remaining terms are as in Eq.~(\ref{eq:F_red}), where $\frac{\gamma}{2}(\partial_x h)^2$ can be obtained from $\gamma\xi$ in the long-wave limit $\vert\partial_xh\vert\ll1$. Minimization of  $F_s^\text{meso}$ with respect to $h$ and $b$ yields
\begin{align}
p &= \partial_h f -\gamma\kappa, \label{app:eq:meso_p}\\
\mu_0 &= \partial_b g_b+\partial_b f-\sigma_b\Delta b.\label{app:eq:meso_mu}
\end{align}
Equations (\ref{app:eq:meso_p}) and (\ref{app:eq:meso_mu}) are the Euler-Lagrange equations of the variational problem. The associated first integral is
\begin{equation}
E = -\frac{\gamma}{\xi}+\frac{\sigma_b}{2}(\partial_x b)^2-f-g_b+\mu_0b+ph,\label{app:eq:first_int}
\end{equation}
where $E$ is a constant. 

Following Ref.~\citep{TSTJ2018l}, we consider Eqs.~(\ref{app:eq:meso_p})-(\ref{app:eq:first_int}) in two different regions, namely in the adsorption layer far away from the drop and in a wedge region of the drop that is defined by a vanishing curvature and where the film thickness is large.  Specifically, we have in the adsorption layer $h\rightarrow h_a$, $b\rightarrow b_a$,~i.e., both fields are spatially uniform. In the wedge region, we have $\kappa\rightarrow 0$ and  $f, \partial_h f, \partial_b f\rightarrow 0 $. We also have $b\rightarrow b_w$, where $b_w$ is the spatially uniform density of $B$ in this region, and $1/\xi\rightarrow \cos\theta_e$ (that is, the slope of the wedge is directly related to the equilibrium contact angle). Note that $b_w$ may be identified with $b_d$ in the macroscopic picture. In the adsorption layer, Eqs.~(\ref{app:eq:meso_p})-(\ref{app:eq:first_int}) become
\begin{align}
p&=\partial_h f(h_a, b_a),\\
\mu_0 &= \partial_bg(b_a)+\partial_bf(h_a,b_a),\label{app:eq:mu_0_adsorption}  \\
E&=-\gamma-f(h_a, b_a)-g_b(b_a)+\mu_0 b_a+ph_a.
\end{align}
In the wedge region, we have
\begin{align}
p&=0, \label{app:eq:p_zero}\\
\mu_0 &=\partial_bg_b(b_w), \label{app:eq:mu_0_wedge} \\
E&= -\gamma\cos\theta_e-g_b(b_w)+\mu_0b_w. \label{app:eq:first_int_wedge}
\end{align}
Note that in Eq.~(\ref{app:eq:first_int_wedge}) we have already used Eq.~(\ref{app:eq:p_zero}). Since $p,\mu_0$ and $E$ are constants, we obtain by comparing wedge and adsorption layer region
\begin{align}
0 &= \partial_h f(h_a, b_a), \\
\partial_b g_b(b_w) &= \partial_bg_b(b_a)+\partial_bf(h_a, b_a), \\
\gamma\cos\theta_e +g_b(b_w) -\mu_0b_w &= \gamma+f(h_a,b_a)+g_b(b_a)-\mu_0b_a.\label{app:eq:first_int_comparison}
\end{align}
Rearranging yields the mesoscopic Young law
\begin{equation}
\gamma(\cos\theta_e-1)=g_b(b_a)-\mu_0b_a+f(h_a,b_a)-\left[g_b(b_w)-\mu_0b_w\right]\label{app:eq:young_meso}
\end{equation}
By comparison with Eq.~(\ref{app:eq:young_macro}), the right hand side of (\ref{app:eq:young_meso}) is the spreading coefficient. Further, we identify expressions for the (macroscopic) interfacial tensions,
\begin{align}
\gamma_{sl} &=g_b(b_w)-\mu_0b_w,\label{app:eq:gamma_sl_meso}\\
\gamma_{sg} &= g_b(b_a)-\mu_0b_a+f(h_a,b_a)+\gamma.\label{app:eq:gamma_sg_meso}
\end{align}
We see from Eqs.~(\ref{app:eq:gamma_sl_meso})-(\ref{app:eq:gamma_sg_meso}) that the mesoscopic expression for $\gamma_{sl}$ is identical to the macrosopic one (\ref{app:eq:gamma_sl_macro}) when $b_w$ is identified with $b_d$. By contrast, the solid-gas interfacial tension $\gamma_{sg}$ is represented mesoscopically by the sum of the energetic contributions of the solid-liquid interface, the adsorption layer and the liquid-gas interface, respectively given by $g_b(b_a)-\mu_0b_a$, $f(h_a,b_a)$ and $\gamma$.

We next show that for the choice $f(h, b) = A_h\left(1+\lambda\frac{b}{b_0}\right)\left(-\frac{1}{2h^2}+\frac{h_a^3}{h^5}\right)$ and with $g_b(b) = \gamma_{sl}^0+ k_\mathrm{B}T b\left[\ln b/b_0 -1\right]$ as in the main text, one has $S<0$ corresponding to partial wetting. More generally, we consider wetting energies of the form $f(h,b) = k_1(1+k_2b)\hat{f}(h)$, where $\hat{f}$ is a function with a single global minimum at $h=h_a$ with $\hat{f}(h_a)<0$ and $k_1, k_2>0$. Using Eqs.~(\ref{app:eq:mu_0_adsorption}), (\ref{app:eq:mu_0_wedge}), (\ref{app:eq:young_meso}) and the relation $g_b=\gamma^0_{sl}+b\left[\partial_bg_b-k_\mathrm{B}T\right]$ (which is only valid for purely entropic $g_b$), we then have for the spreading coefficient
\begin{equation}
\begin{split}
S&=b_a\left[\mu_0-k_1k_2\hat{f}(h_a)-k_\mathrm{B}T\right]-b_w\left[\mu_0-k_\mathrm{B}T\right]-\mu_0(b_a-b_w)+k_1(1+k_2b_a)\hat{f}(h_a),\\
&=-k_\mathrm{B}T(b_a-b_w)+k_1\hat{f}(h_a) <0.
\end{split}
\end{equation}
The final inequality follows from $b_a>b_w$, which is obtained from (\ref{app:eq:mu_0_adsorption}) and (\ref{app:eq:mu_0_wedge}), using $\hat{f}(h_a)<0$ and $k_1, k_2>0$.

\section{Nondimensionalization}\label{app:nondim}
Here, we briefly describe nondimensionalization of Eqs.~(\ref{eq:dt_h_reduced})-(\ref{eq:R_bc_reduced}). To this end, we introduce the scales 
\begin{equation}
(x, y) = L (\tilde{x}, \tilde{y}), \hspace{0.5cm} h= l \tilde{h}, \hspace{0.5cm} t = \tau\tilde{t}, \hspace{0.5cm} b = b_0\tilde{b},
\end{equation}
with the scaling factors 
\begin{equation}
L = l\sqrt{\frac{\gamma}{k_b T b_0}}, \hspace{0.5cm} l = h_a, \hspace{0.5cm} \tau = \frac{L^2\eta}{h_a k_b T b_0}.
\end{equation}
Dimensionless quantities are denoted by tildes. The dimensionless time evolution equations are then given by
\begin{align}
\partial_{\tilde{t}} \tilde{h} &= \tilde{\vec{\nabla}}\cdot\left[\frac{\tilde{h}^3}{3}\tilde{\vec{\nabla}} \tilde{p}\right],\label{app:eq:dt_h_reduced_nondimensional}\\
\partial_{\tilde{t}} \tilde{b} &= \tilde{\vec{\nabla}}\cdot\left[\tilde{D}_b \tilde{b} \tilde{\vec{\nabla}}\tilde{\mu}_b\right]+\tilde{R}_{ab}-\tilde{R}_{bc},\label{app:eq:dt_b_reduced_nondimensional}
\end{align}
with the nondimensional pressure and chemical potential given as 
\begin{align}
\tilde{p} &= W\cdot\left(1+\lambda \tilde{b}\right)\cdot\left(\frac{1}{\tilde{h}^3}-\frac{1}{\tilde{h}^6}\right)-\tilde{\vec{\nabla}}^2\tilde{h}, \\
\tilde{\mu}_b &= \ln\tilde{b}+\lambda W \left(-\frac{1}{2\tilde{h}^2}+\frac{1}{5\tilde{h}^5}\right)-\tilde{\sigma}_b\tilde{\vec{\nabla}}^2\tilde{b}.
\end{align}
The nondimensional reactive currents are 
\begin{align}
\tilde{R}_{ab} &= \tilde{\bar{r}}_1(\tilde{h})\left[\exp\left(\tilde{\mu}_a\right)-\exp\left(\tilde{\mu}_{b}\right)\right], \\
\tilde{R}_{bc} &= \tilde{\bar{r}}_2(\tilde{h})\left[\exp\left(\tilde{\mu}_b\right)-\exp\left(\tilde{\mu}_{c}\right)\right],
\end{align}
where the rate functions are given by
\begin{align}
\tilde{\bar{r}}_1(\tilde{h}) &=  \frac{\tilde{r}_{1}}{2}\left[1+\tanh\left(\frac{\tilde{h}-\tilde{h}_0}{\Delta \tilde{h}}\right)\right], \\
\tilde{\bar{r}}_2(\tilde{h}) &=  \frac{\tilde{r}_{2}}{2}\left[1-\tanh\left(\frac{\tilde{h}-\tilde{h}_0}{\Delta\tilde{h}}\right)\right]. \label{app:eq:stepfunc_2_nondimensional}
\end{align}
The remaining dimensionless quantities are 
\begin{align}
W &= \frac{A_h}{h_a^2 b_0 k_\mathrm{B} T}, \\
\tilde{\sigma}_b &= \frac{b_0}{L^2 k_\mathrm{B} T}\sigma_b,\\
\tilde{D}_b &= \frac{\tau k_\mathrm{B} T}{L^2}D_b, \\
\tilde{r}_{1,2} &= \frac{\tau}{b_0}r_{1,2},\\
\tilde{\mu}_{a,c} &= \frac{1}{k_\mathrm{B} T} \mu_{a,c}, \\
\tilde{h}_0 &= \frac{1}{h_a}h_0, \\
\Delta\tilde{h} &= \frac{1}{h_a}\Delta h.
\end{align}
In the main text, tildes denoting nondimensional quantities are omitted.

\end{document}